\documentclass [aps,prb,twocolumn,amssymb,showpacs,superscriptaddress,reprint,citeautoscript]{revtex4-1}

\usepackage{lineno}
\usepackage[pdftex]{graphicx}
\usepackage{float}
\usepackage{bm}

\begin{document}

\title{Angular-dependent upper critical field of overdoped Ba(Fe$_{1-x}$Ni$_x$)$_2$As$_2$ }%

\author{J.~Murphy}
%\email{jmurph@iastate.edu}
\affiliation{Ames Laboratory US DOE, Ames, Iowa 50011, USA}
\affiliation{Department of Physics and Astronomy, Iowa State University, Ames, Iowa 50011,
USA }

\author{M.~A.~Tanatar}
%\email{tanatar@ameslab.gov}
\affiliation{Ames Laboratory US DOE, Ames, Iowa 50011, USA}
\affiliation{Department of Physics and Astronomy, Iowa State University, Ames, Iowa 50011,
USA }

\author{D.~Graf}
%\email{graf@magnet.fsu.edu}
\affiliation{National High Magnetic Field Laboratory, Florida State university, Tallahassee, Florida 32310, USA}

\author{J.~S.~Brooks}
%\email{brooks@magnet.fsu.edu}
\affiliation{National High Magnetic Field Laboratory, Florida State university, Tallahassee, Florida 32310, USA}

\author{S.~L.~ Bud'ko}
%\email{budko@ameslab.gov}
\affiliation{Ames Laboratory US DOE, Ames, Iowa 50011, USA}
\affiliation{Department of Physics and Astronomy, Iowa State University, Ames, Iowa 50011,
USA }

\author{P.~C.~Canfield}
%\email{canfield@ameslab.gov}
\affiliation{Ames Laboratory US DOE, Ames, Iowa 50011, USA}
\affiliation{Department of Physics and Astronomy, Iowa State University, Ames, Iowa 50011,
USA }

\author{V.~G.~Kogan}
%\email{kogan@ameslab.gov}
\affiliation{Ames Laboratory US DOE, Ames, Iowa 50011, USA}

\author{R.~Prozorov}
\email[Corresponding author: ]{prozorov@ameslab.gov}
\affiliation{Ames Laboratory US DOE, Ames, Iowa 50011, USA}
\affiliation{Department of Physics and Astronomy, Iowa State University, Ames, Iowa 50011,
USA }

\date{27 January 2013}

\begin{abstract}
In-plane resistivity measurements as a function of temperature, magnetic field and its orientation with respect to the crystallographic $ab-$plane were used to study the upper critical field, $H_{c2}$, of two overdoped compositions of the iron-based superconductor Ba(Fe$_{1-x}$Ni$_x$)$_2$As$_2$, $x=$0.054 and $x$=0.072. Measurements were performed using precise alignment (with accuracy less than 0.1$^o$) of magnetic field with respect to the Fe-As-plane. The dependence of the $H_{c2}$ on angle $\theta$ between the field and the $ab$-plane was measured in isothermal conditions in a broad temperature range. We found that the shape of $H_{c2}(\theta)$, clearly deviates from Ginzburg-Landau functional form.

\end{abstract}

\maketitle

\section{Introduction}

The theory of the upper critical field in superconductors was essentially developed by the mid-60s \cite{HW,WHH,CC}, with the linear $H_{c2}(T)$ behavior close to $T_c$ described by the anisotropic Ginzburg-Landau theory \cite{Gorkov,Tilley}, leading in tetragonal crystals to dependence of $H_{c2}$ on angle $\theta$ with respect to the $ab$-plane:

\begin{equation}
H_{c2}(\theta)=\frac{H_{c2,ab}}{\sqrt{(\gamma_H ^2 -1) \sin^2\theta + 1 }}\,,\qquad \gamma_H=\frac{H_{c2,ab}}{H_{c2,c}}\,.
\label{Hc2-theta}
\end{equation}

\noindent

Renewed interest to studies of the upper critical field was brought by the discovery of materials with potentially unconventional pairing mechanisms, heavy fermion, organic and especially cuprate superconductors, with the well documented $d$-symmetry of the superconducting order parameter in the latter case \cite{dwave}. It was quickly recognized that the angular dependence of the upper critical field, in particular on the angle $\phi$ in the basal plane of the tetragonal crystals, can be used to probe the anisotropy of the superconducting order parameter \cite{Gorkov1984}. These ideas were further developed for unconventional superconductors with various exotic order parameters in a series of papers by K.~Maki and co-workers \cite{Makidwave,Makids,Makipwave,Makifwave,Makianisotropicswave}. They stimulated a series of experimental studies, in particular in Sr$_2$RuO$_4$ \cite{Mao,TanatarSROPRB,Deguchi}, heavy fermion \cite{Weickert} and organic superconductors \cite{Lee,Kawasaki,Kamiya,Kovalev,TanatarlambdaBETS}.

Although these developments were correctly catching the importance of the anisotropy of the superconducting gap structure for the angular-dependent $H_{c2}$, the theoretical data analysis was oversimplified by assumption of simple cylindrical or spherical Fermi surface shapes. The importance of the Fermi surface topology for the anisotropy of the $H_{c2}$ was brought to focus by the discovery of pronounced multi-band effects in superconductivity of MgB$_2$ \cite{MgB2discovery,MgB2paranoya1,MgB2multiband}. The upper critical field of this layered material is anisotropic and can be varied significantly by carbon and aluminum doping \cite{MgB2carbondoping,MgB2Al,Angst}, controlling the mean free path of the carriers and changing the inter-band coupling. Theoretical modeling explicitly included these effects into consideration and produced the angular dependence of the $H_{c2}(\theta)$ which was virtually identical to the dependence of Eq.~\ref{Hc2-theta} \cite{MgB2theta}.

Recent developments in understanding of the upper critical field were greatly stimulated by the discovery in 2008 of the layered FeAs superconductors \cite{Hosono}, which opened up a new avenue in high $T_c$ research.
The upper critical fields of iron pnictides are very high \cite{NatureNHMFL}, and besides the potential for high field applications \cite{pnictides_applications}, this brings the possibility of the paramagnetic effects at low temperatures \cite{LiFeAs_Kyuil}.

Since Fe-As layers form a main building block of all iron-based superconductors \cite{Paglione,Johnston,CanfieldBudko,Stewart}, these compounds show anisotropy of the electronic structure, reflected in the anisotropy $\gamma_H$ of the upper critical field. Unlike signature layered materials, organic superconductors \cite{Ishiguro} and cuprates \cite{cuprates}, the values of $\gamma_H$ in iron arsedines are small, for most compounds in the range 2 to 4 \cite{Wenaniz,Hc2a1,altrawneh,NiNiCo,Hc2a2,Hc2a3,Hc2a4,anisotropy,NiNiNiparanoya,NiNiNi}, with few exceptions \cite{WelpSm,NiNi10-3-8,Balicas1111} were values as high as 7 to 10 are found, see Ref.~\onlinecite{Gurevich} for review. Furthermore $\gamma _H$ in iron arsenides shows notable temperature dependence, presumably reflecting multi-band effects \cite{anisotropy}.
Low anisotropy values suggest that at least close to the critical in zero-field, $T_c (H=0)$,  we are dealing with orbital $H_{c2}$.

The detailed study of the anisotropy of the upper critical field along principal $a$ and $c$ directions as a function of doping was undertaken in the series of electron-doped compounds Ba(Fe$_{1-x}$Co$_x$)$_2$As$_2$ (BaCo122 in the following) \cite{NiNiCo}. It was found that $\gamma_H$ changes significantly between under-doped $x<0.08$ and overdoped regions of the phase diagram, presumably reflecting change of the electronic structure. Comparison with optimally doped Ba(Fe$_{1-x}$Ni$_x$)$_2$As$_2$ (BaNi122 in the following), $x$=0.046 \cite{NiNiNiparanoya,NiNiNi} suggested that the anisotropy is not particularly sensitive to the amount of disorder introduced by dopant $x$.  

On the other hand, several studies pointed out complex superconducting gap structure of iron arsenides, both in BaCo122 \cite{Tanatarthermalcond,Reidcaxis,heatcapacityBudko,Tanigaki,Ronning,hcStewart} and analogous BaNi122 \cite{Martin}. For both these materials, it was suggested that notable superconducting gap modulations along $c$-axis should be characteristic of the overdoped regime.
Since this gap variation might be reflected in the angular dependence of the $H_{c2}(\theta)$, we decided to perform systematic study of these compositions. To our knowledge, there was only one study addressing angular variation of $H_{c2}$ in iron-based superconductors \cite{Singleton}, providing no high angular resolution data.

In this paper we perform a detailed characterization of the upper critical field in overdoped BaNi122 as a function of temperature and field direction. Our main experimental finding is a clear deviation of the angular dependence from Eq.~\ref{Hc2-theta},  particularly strong for the directions of magnetic field close to $H \parallel c$, where orbital effects should be the strongest. 

%%%%%%%%%%%%%%%%%%%%%%%%%%%%%%%%%%%%%%Experimental

\section{Experimental}
\begin{figure}[tb]
\begin{center}
\includegraphics[width=0.90\linewidth]{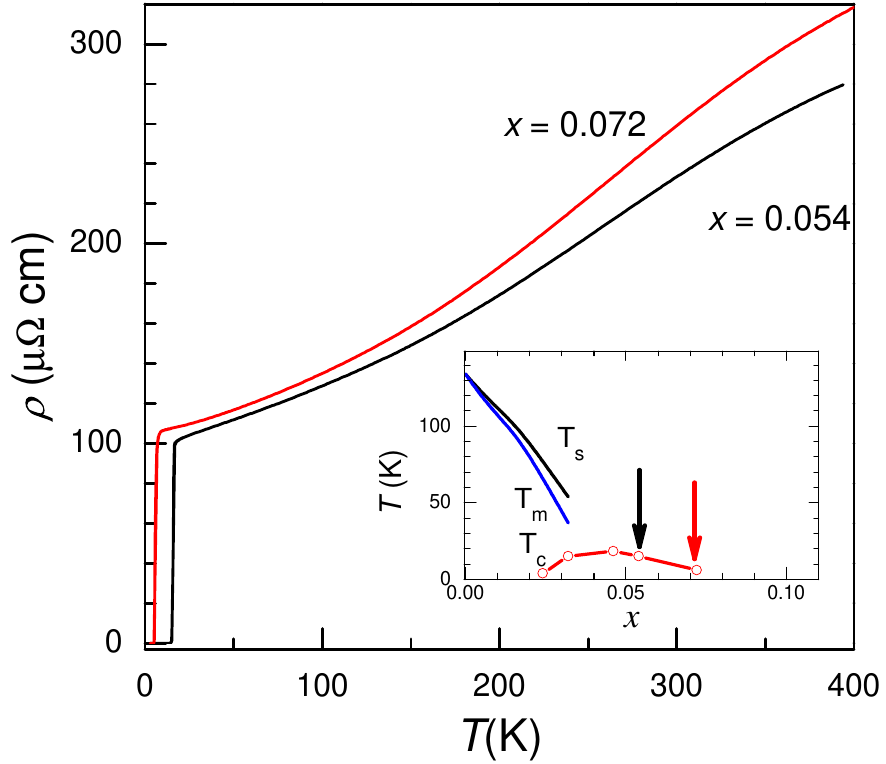}
\end{center}
\caption{(Color Online) Temperature-dependent resistivity of two samples of BaFe$_{1-x}$Ni$_x$As$_2$ used in this study, with $x$=0.054 (slightly overdoped) and $x$=0.072 (strongly overdoped), with doping level indicated with arrows with respect to temperature-doping phase diagram of BaNi122 after Ref.~\onlinecite{NiNiNiparanoya,NiNiNi} shown in the inset. Note pronounced curvature of the $\rho (T)$ for $T>T_c$, typical of overdoped compositions \cite{NDL}. Sample resistivity value is defined with accuracy of about 20\% due to uncertainty of geometric factors, see Refs.~\onlinecite{anisotropy,pseudogap} for details.
}%
\label{resistivity}
\end{figure}

\begin{figure}[tb]
\begin{center}
\includegraphics[width=0.90\linewidth]{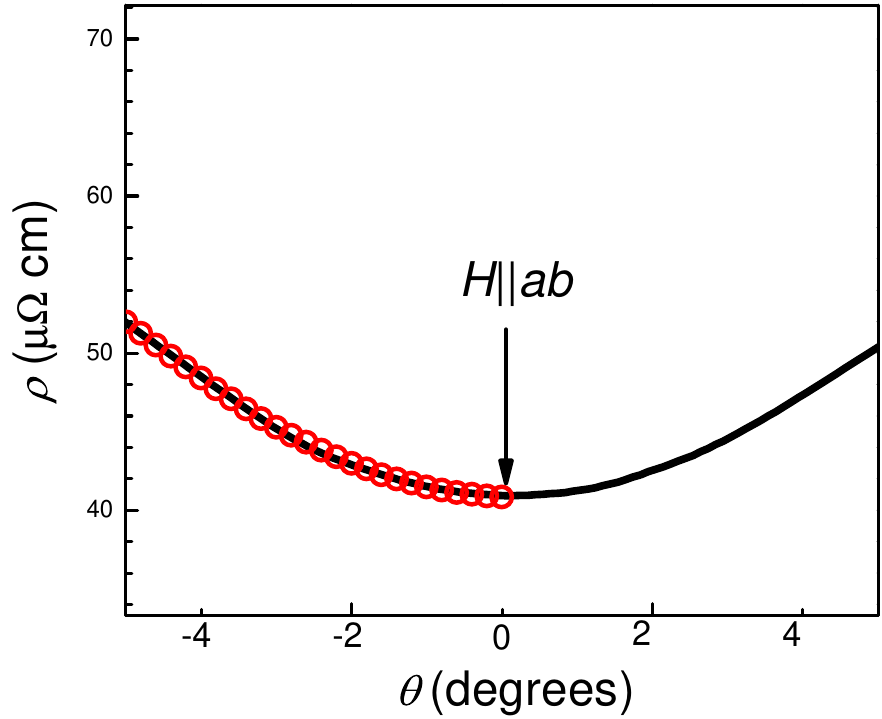}
\includegraphics[width=0.90\linewidth]{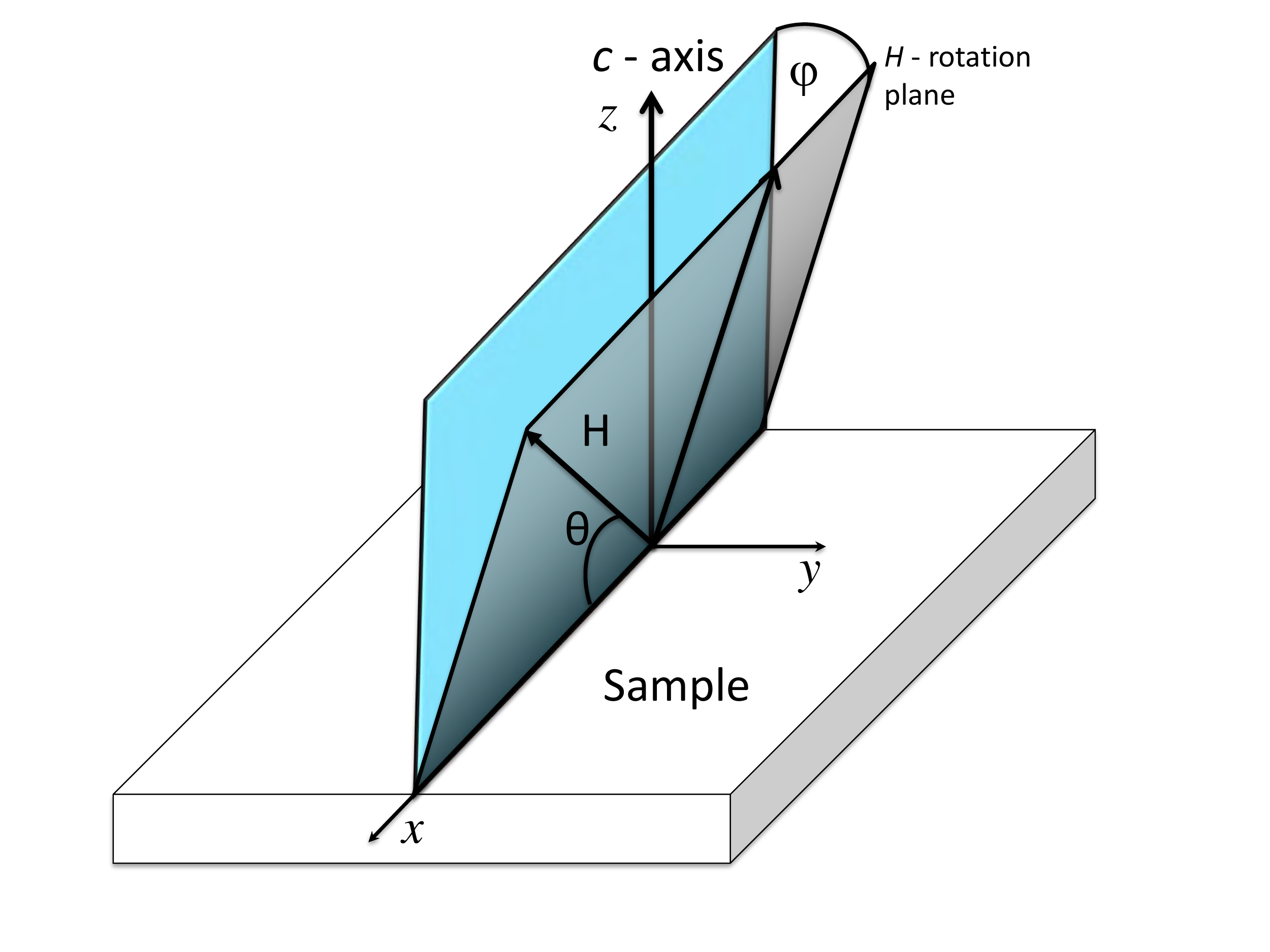}
\end{center}
\caption{(Color Online) During experiments in single axis rotation system of 35~T magnet, the direction of magnetic field was aligned parallel to the conducting plane by resistivity measurements in field $H$ slightly lower than $H_{c2 \parallel}$, in which sample resistance shows strong angular dependence, black line in the top panel. The curve was measured in one-sided motion of the rotator to avoid backlash, with deep minimum corresponding to $H \parallel ab$ or $\theta$=0 condition. The red open symbols show alignment measurements, taken in a second angular sweep of the same rotation direction, and stopped at $\theta$=0. $H$ and $T$ sweeps were used to determine the phase diagrams in $H \parallel ab$ condition, and then magnetic field angle $\theta$ with respect to the plane was changed by continuing rotation of the sample in the same direction as during alignment. Because the orientation of the sample in the third direction, perpendicular to the rotation plane, was set by eye there may exist non-zero angle $\varphi$ between the field-rotation plane and the plane of the normal to the sample. In most cases this angle should be less than 5$^o$.
}%
\label{scheme}
\end{figure}

Single crystals of Ba(Fe$_{1-x}$Ni$_x$)$_2$As$_2$ were grown from FeAs flux using a starting load of metallic Ba, FeAs and NiAs, as described in detail elsewhere \cite{NiNiNi}. Crystals were thick platelets with large faces corresponding to the tetragonal (001) plane.
Actual composition of the crystals was determined using WDS x-ray electron probe microanalysis \cite{NiNiNi}. The two compositions studied were on the overdoped side of the phase diagram, slightly overdoped $x$=0.054 ($T_c$=16~K) and strongly overdoped $x$=0.072 ($T_c$=7.5~K), whereas maximum $T_c$=19~K is achieved at optimal doping, $x_{opt}$=0.046 \cite{NiNiNiparanoya,NiNiNi}, see doping phase diagram in inset in Fig.~\ref{resistivity}.

Samples for in-plane resistivity, $\rho$, measurements were cleaved with a razor blade into rectangular strips with typical dimensions, $2\times(0.1-0.3)\times(0.03-0.1)$ mm$^3$ and the long side corresponding to tetragonal $a$-axis. All sample dimensions were measured with an accuracy of about 10\%. Contacts to the samples were made by attaching silver wires using ultrapure tin, resulting in an ultra low contact resistance (less than 10 $\mu \Omega$) \cite{SUST}. Resistivity measurements were made using a standard four-probe technique, producing the $\rho(T)$ curves as shown in Fig.~\ref{resistivity}. After initial preparation, samples were characterized in PPMS system, and then glued by GE-varnish to a plastic platform, fitting single axis rotator of the 35~T DC magnet in National High Magnetic Field Laboratory in Tallahassee, Florida. Sample resistance was checked after mounting and found to be identical to the initial value. High-field measurements were made in He-cryostat with variable temperature control inset (VTI) allowing for temperatures down to 1.5~K.

The stepping motor driven rotator enabled {\it in situ} rotation with 0.05$^o$ resolution around a horizontal axis in single axis rotation system of vertical 35~T magnetic field. During this rotation the direction of magnetic field with respect to the crystal stays in a plane of rotation, see Fig.~\ref{scheme}. We can precisely align the direction of the magnetic field parallel to the sample plane within the rotation plane, defined as $\theta=$0, using angular dependence of resistivity, measured in magnetic field slightly below $H_{c2 \parallel}$. This alignment is illustrated in Fig.~\ref{scheme}. In an ideal case of the second sample axis coinciding with the rotation axis, field-rotation plane should contain $c$-axis of the sample. There may have been a non-zero uncontrolled angle $\varphi$ between the field-rotation plane and the plane of the normal to the sample, see Fig.~\ref{scheme}. We estimate that $\varphi <5^\circ$.

\section{Results}

%%%%%%%%%%%%%%%%%%%%%%%%%%%%%%%%%H-T phase diagrams

\begin{figure}[tb]
\begin{center}
\includegraphics[width=0.90\linewidth]{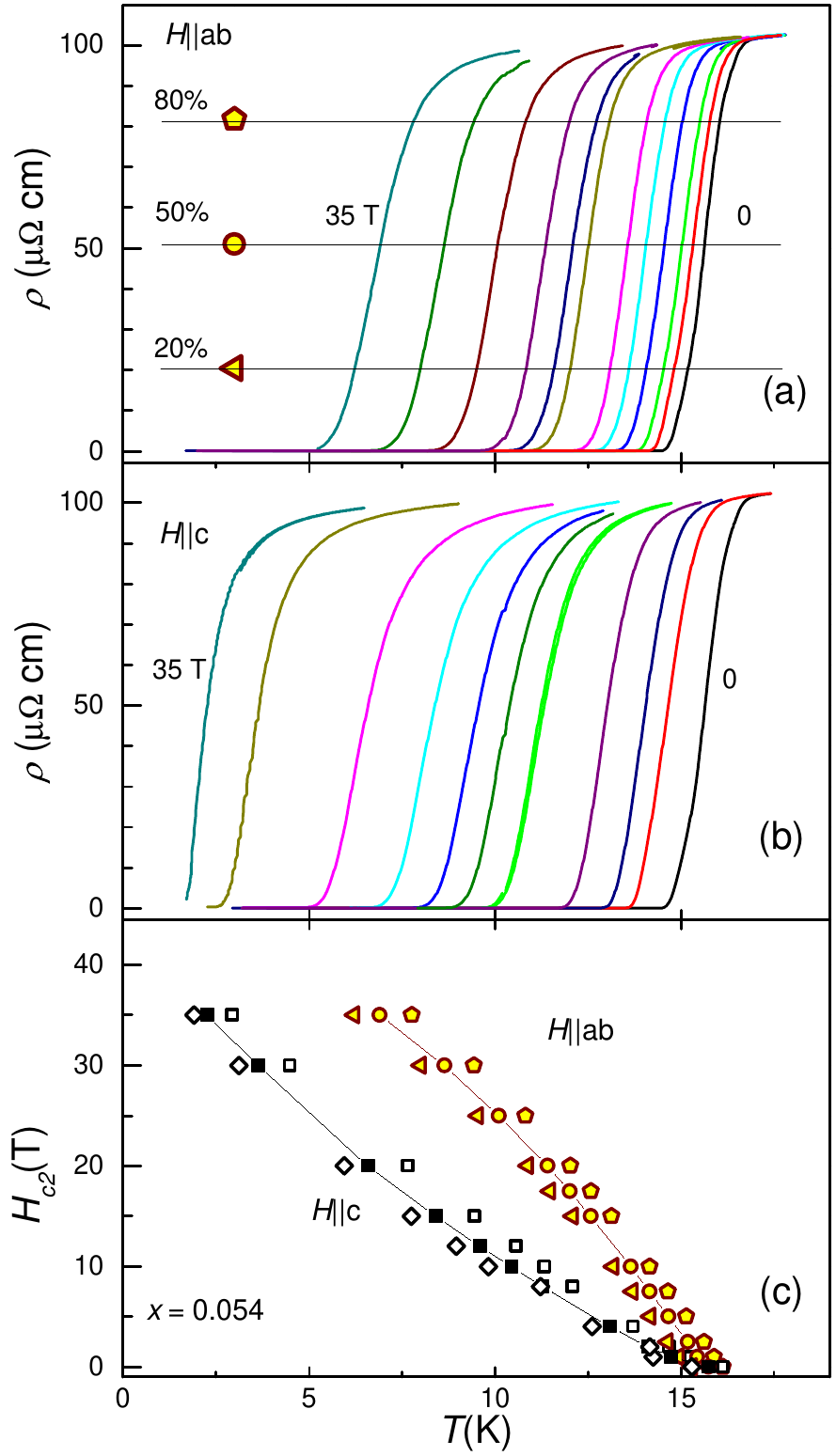}
\end{center}
\caption{(Color Online) In-plane resistivity $\rho_{a}$ vs. temperature for slightly overdoped Ba(Fe$_{1-x}$Ni$_{x}$)$_2$As$_2$, $x$=0.054 in magnetic fields (a) parallel to the conducting $ab$ plane (right to left, 0~T, 1~T, 2.5~T, 5~T, 7.5~T, 10~T, 15~T, 17.5~T, 20~T, 25~T, 30~T, 35~T); (b) parallel to the $c$-axis (right to left, 0~T, 1~T, 2~T, 4~T, 8~T, 10~T, 12~T, 15~T, 20~T, 30~T, 35~T). Lines indicate 20, 50, and 80 \% of the resistivity value immediately above the superconducting transition. Bottom panel (c) shows $H_{c2}(T)$ phase diagrams for both directions of magnetic field.
}%
\label{phasedslight}
\end{figure}

\begin{figure}[tb]
\begin{center}
\includegraphics[width=0.90\linewidth]{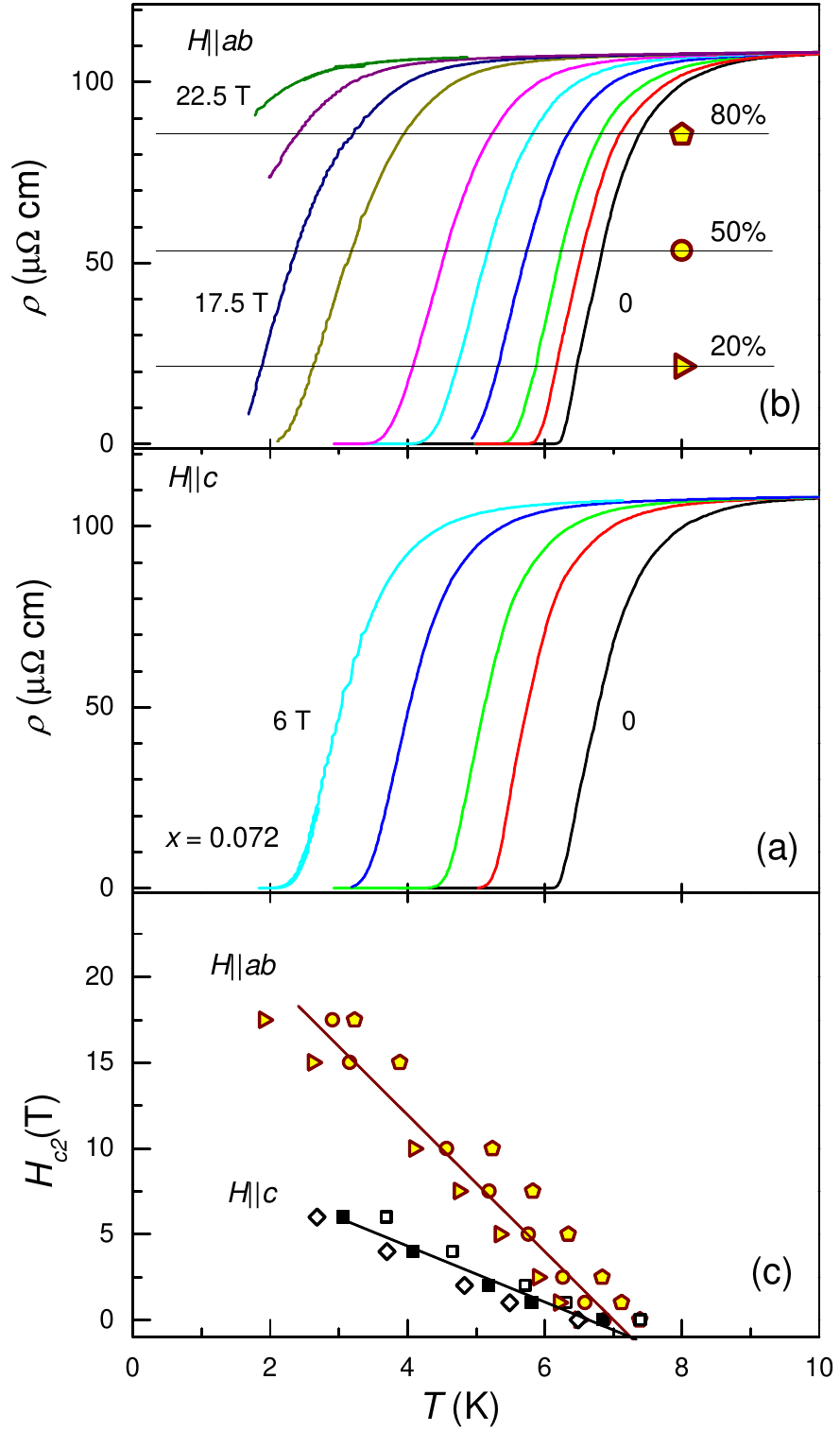}
\end{center}
\caption{(Color Online) In-plane resistivity $\rho_{a}$ vs. temperature for heavily overdoped Ba(Fe$_{1-x}$Ni$_{x}$)$_2$As$_2$, $x$=0.072 in magnetic fields (a) parallel to the conducting $ab$ plane (right to left, 0~T, 1~T, 2.5~T, 5~T, 7.5~T, 10~T, 15~T, 17.5~T, 20~T, 22.5~T); (b) parallel to the $c$-axis (right to left, 0~T, 1~T, 2~T, 4~T, 6~T). Lines indicate 20, 50, and 80 \% of the resistivity value immediately above the superconducting transition. Bottom panel (c) shows $H_{c2}(T)$ phase diagrams for both directions of magnetic field.
}%
\label{phasedstrong}
\end{figure}

In Figs.~\ref{phasedslight} and ~\ref{phasedstrong} we show raw $\rho(T)$ data for a set of magnetic fields aligned approximately along $c$-axis ($\theta=$90$^o$, top panels) and precisely along the conducting plane ($\theta=$0$^o$, bottom panels), for BaNi122 samples with $x$=0.054 and $x$=0.072, respectively. We show also the lines corresponding to 20, 50 and 80\% of resistivity value immediately above the transition, $\rho (T_c)$, used as criteria to determine the transition temperature as a function of magnetic field and construct the phase diagrams, bottom panels (c) of Figs.~\ref{phasedslight} and \ref{phasedstrong}. The use of these criteria is justified by small variation of the resistive transition width on application of magnetic field, and its independence on the extrapolation, typical problem for onset and offset criteria.

As can be most clearly seen from the bottom panel of Fig.~\ref{phasedslight}, the shapes of the $H_{c2}(T)$ phase diagrams in parallel and perpendicular field orientations share the same features as found in previous studies of other Fe based systems.  The $H_{c2,ab}(T)$ flattens at low temperatures, while $H_{c2,c}(T)$ maintains positive curvature down to the lowest temperatures of our experiment. Both these features are typical for layered materials, see for example Refs.~\onlinecite{Andy,Ishiguro}.

%%%%%%%%%%%%%%%%%%%%%%%%%%%%Angular dependence

\begin{figure}[tb]
\begin{center}
\includegraphics[width=.90\linewidth]{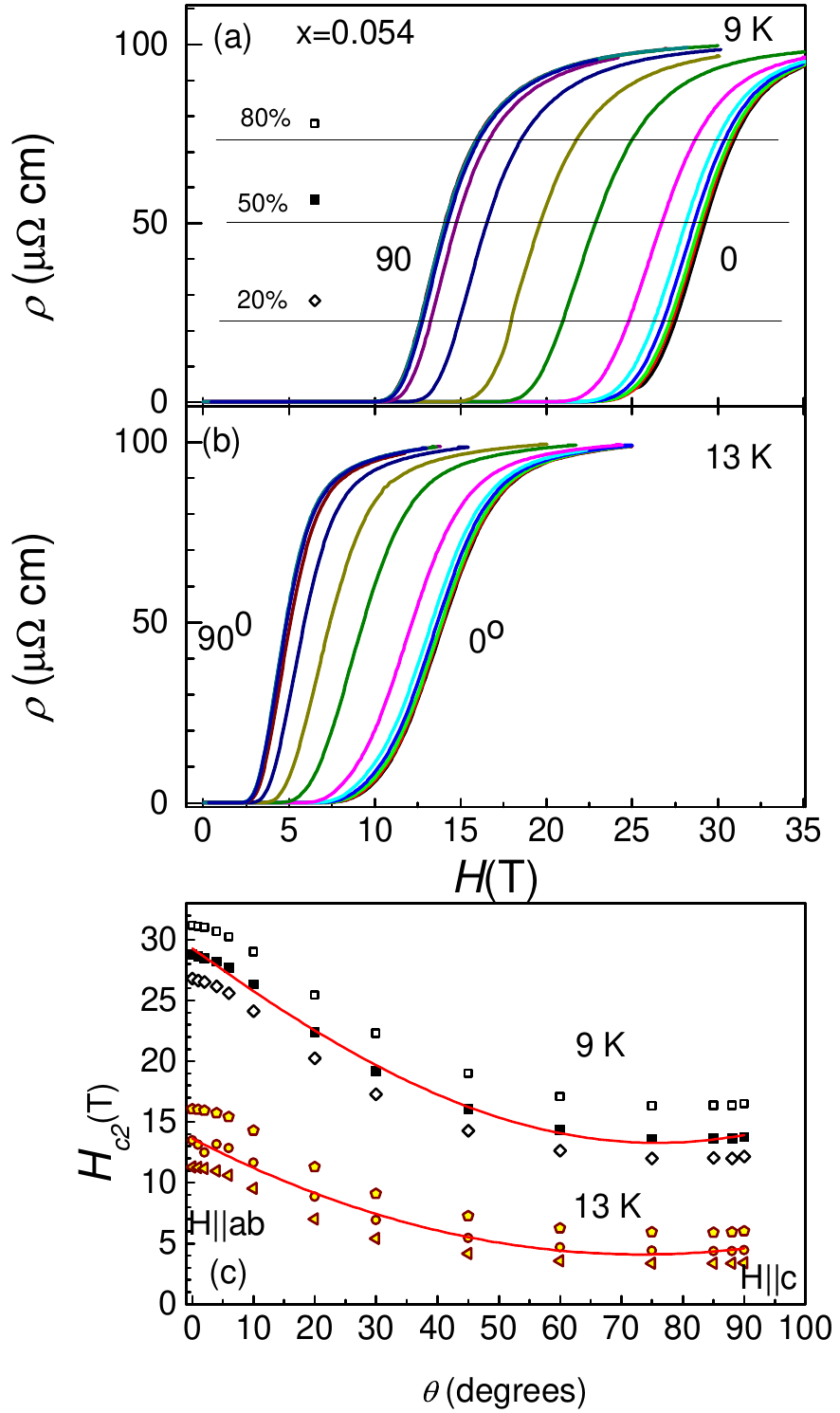}
\end{center}
\caption{(Color Online) Field dependence of in-plane resistivity $\rho (H)$ of slightly overdoped Ba(Fe$_{1-x}$Ni$_{x}$)$_2$As$_2$, $x$=0.054 sample at $T$=13~K (panel (a)) and $T$=9~K (panel (b)) with magnetic field inclination angle $\theta$ as a parameter. (c) Isotherms $H_{c2}(\theta)$, obtained at 9~K and 13~K, using 80\%, 50\% and 20\% criteria. Solid line shows fit to Eq.~\ref{Hc2-theta}.  }%
\label{angleslight}
\end{figure}

\begin{figure}[tb]
\begin{center}
\includegraphics[width=.90\linewidth]{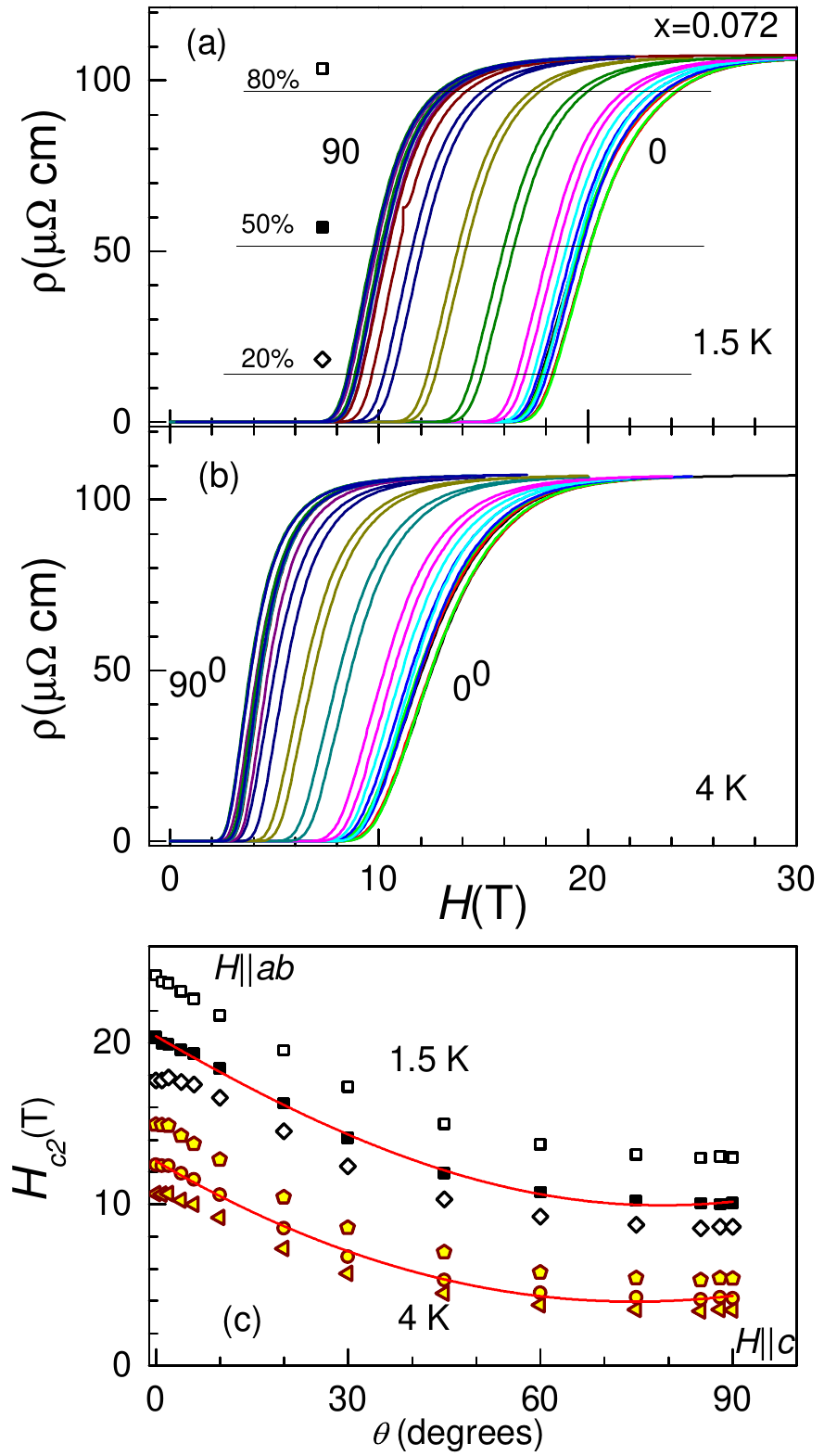}
\end{center}
\caption{(Color Online) Field dependence of in-plane resistivity $\rho (H)$ of strongly overdoped Ba(Fe$_{1-x}$Ni$_{x}$)$_2$As$_2$, $x$=0.072 sample at $T=$1.5~K (panel (a)) and $T$=4~K (panel (b)) with magnetic field inclination angle $\theta$ as a parameter. (c) Isotherms $H_{c2}(\theta)$, obtained 1.5~K and 4~K, using 80\%, 50\% and 20\% criteria. Solid line shows fit to Eq.~\ref{Hc2-theta}.   }%
\label{anglestrong}
\end{figure}

In Figs.~\ref{angleslight} and \ref{anglestrong} we show field dependences of in-plane resistivity taken at fixed temperatures with inclination angle $\theta$ as a parameter for slightly overdoped sample with $x$=0.054 and strongly overdoped sample $x$=0.072, respectively. The data analysis will be presented in the next section.

\section{Discussion}

\subsection{ Angular dependence of $H_{c2}$}

\subsubsection{ Linearization of the $H_{c2}(\theta)$ dependence}

\begin{figure}[tb]
\begin{center}
\includegraphics[width=.90\linewidth]{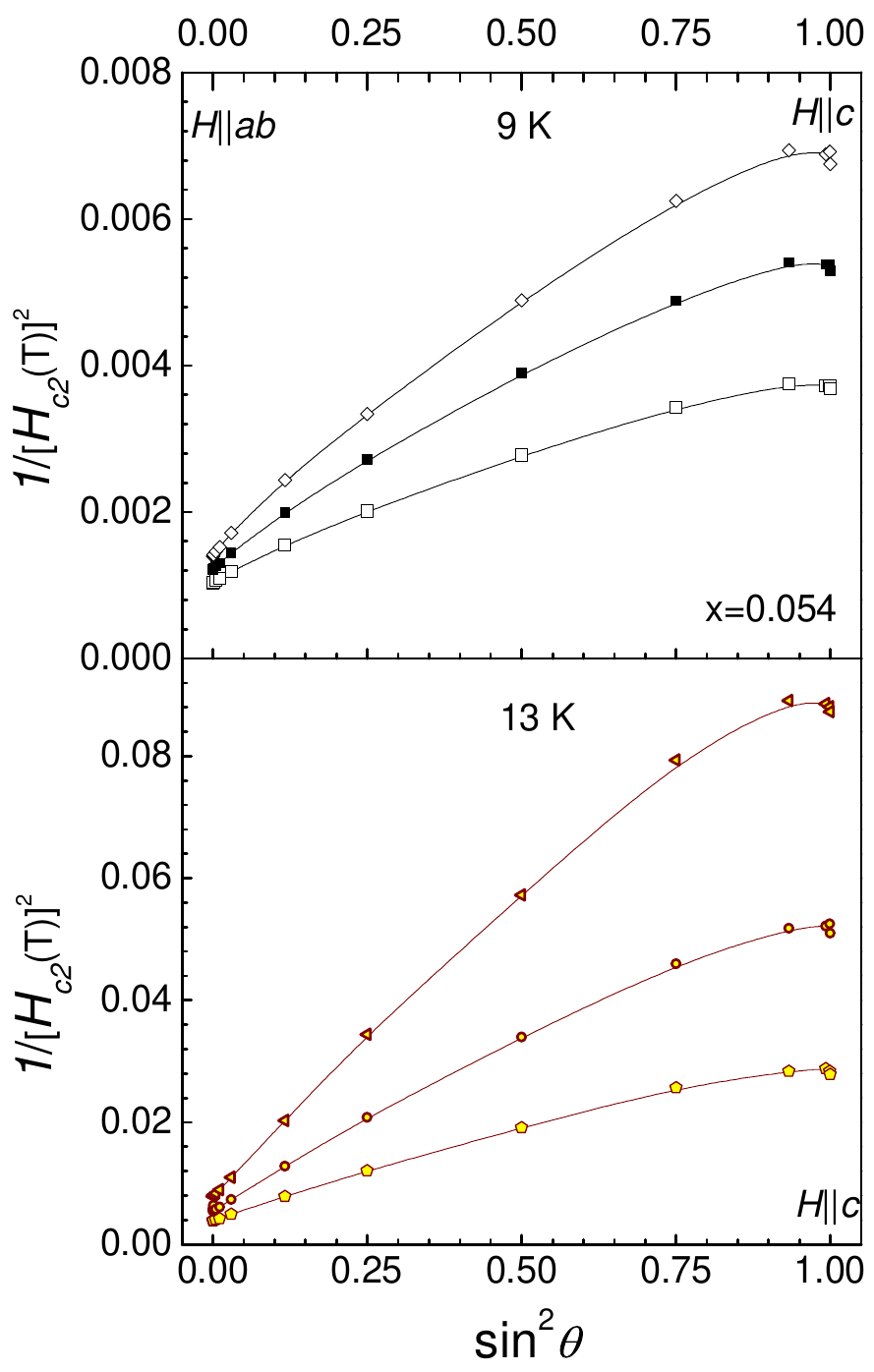}
\end{center}
\caption{(Color Online) Angular dependence $H_{c2}(\theta)$, determined from fixed temperature $\rho(H)$ of Fig.~\ref{angleslight} using 20\%, 50\% and 80\% criteria (top to bottom), for slightly overdoped Ba(Fe$_{1-x}$Ni$_{x}$)$_2$As$_2$, $x$=0.054 at 9~K (top panel) and 13~K (bottom panel). The lines are guides to the eyes. The data are plotted as $H_{c2}^{-2}(\sin^2\theta)$, which according to Eq.~\ref{Hc2-theta} should be a straight line. The lines are guides to the eyes. }%
\label{angleslightanalysis}
\end{figure}

\begin{figure}[tb]
\begin{center}
\includegraphics[width=.90\linewidth]{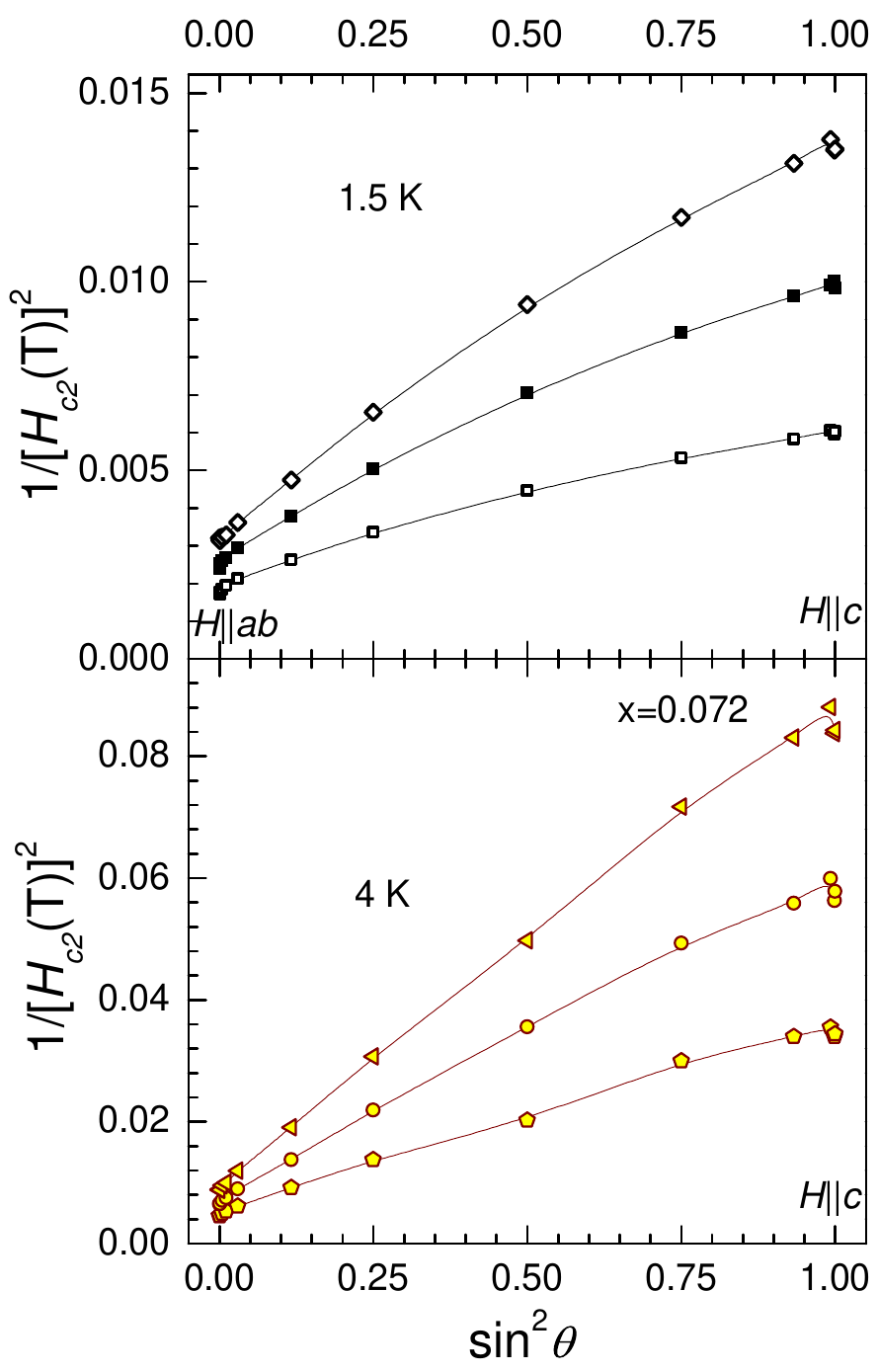}
\end{center}
\caption{(Color Online) Angular dependence $H_{c2}(\theta)$, determined from fixed temperature $\rho(H)$ of Fig.~\ref{anglestrong} using 20\%, 50\% and 80\% criteria (top to bottom), for strongly overdoped Ba(Fe$_{1-x}$Ni$_{x}$)$_2$As$_2$, $x=$0.072 at 1.5~K (top panel) and 4~K (bottom panel). The lines are guides to the eyes.  The data are plotted as $H_{c2}^{-2}(\sin^2\theta)$, which according to Eq.~\ref{Hc2-theta} should be a straight line.  }%
\label{anglestronganalysis}
\end{figure}

%\newpage
%%%%%%%%Analysis for other materials
\begin{figure*}[h!]
\begin{center}
\includegraphics[width=15cm]{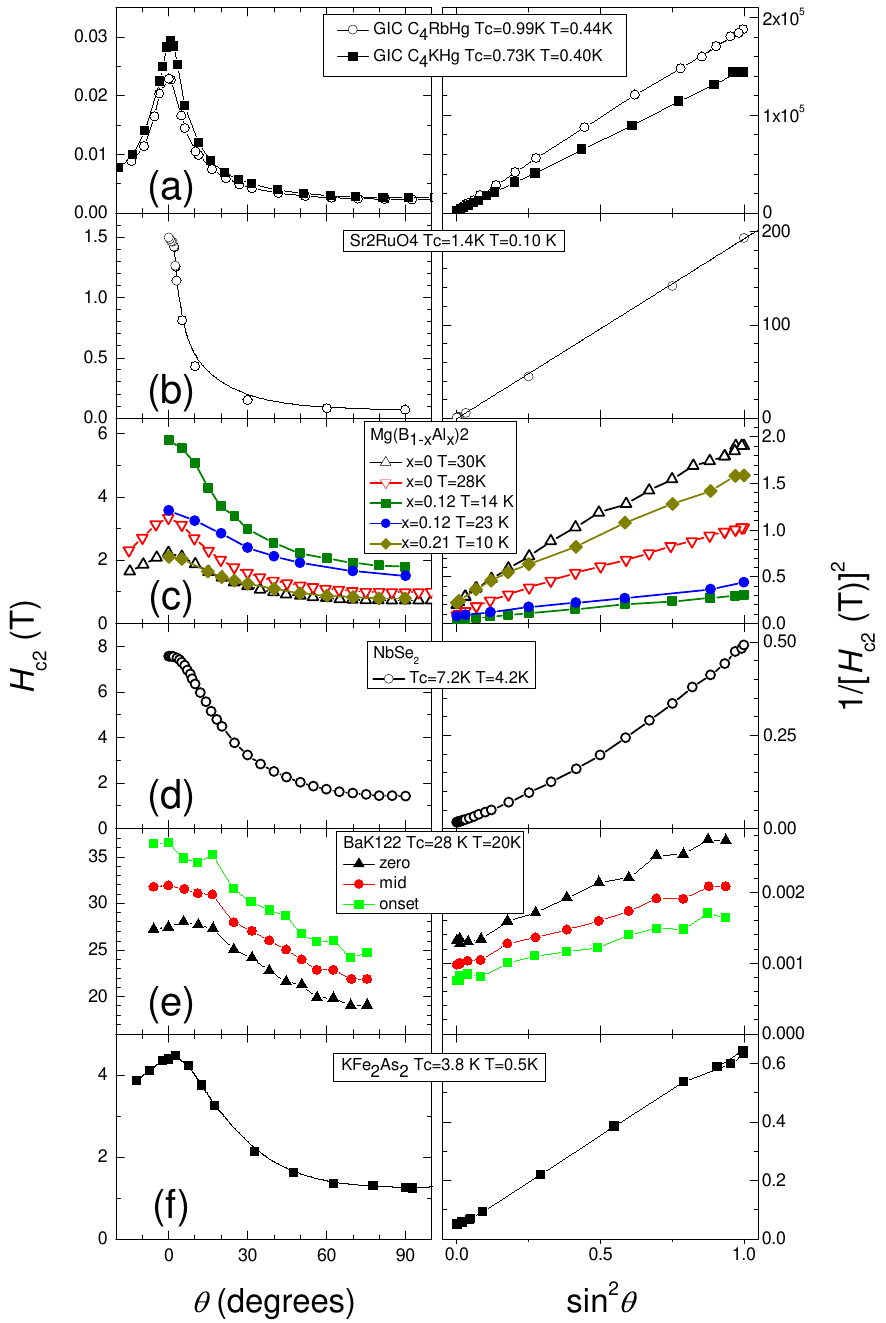}
\end{center}
\caption{(Color Online) Analysis of the isothermal angular dependence of $H_{c2}$ on inclination angle to the highly conducting plane $\theta$, using linearization plot $H_{c2}^{-2}(\sin^2\theta)$.
Left panels show digitized $H_{c2}(\theta)$, right panels plot the same data as $H_{c2}^{-2}(\sin^2\theta)$: (a) Graphite intercalation compounds \cite{Iye} C$_4$RbHg ($T_c$=0.99~K, measurements taken at $T_h$=0.44~K, open circles) and C$_4$KHg ($T_c$=0.73~K, $T_h$=0.40~K, solid squares); (b) Sr$_2$RuO$_4$ ($T_c$=1.43~K, $T_h$=0.10~K, Ref.~\onlinecite{Deguchi}); (c) Mg(B$_{1-x}$Al$_x$)$_2$, Ref.~\onlinecite{MgB2Al}, $x$=0.12
($T_c$=30.8~K, $T_h$=14~K, black solid squares, and $T_h$=23~K, red solid circles) and $x$=0.21 ($T_c$=25.5~K, $T_h$=10~K, blue open circles); (d) NbSe$_2$, Ref.~\onlinecite{NbSe2},
($T_c$= 7.2~K, $T_h$=4.2~K); (e) (Ba$_{1-x}$K$_x$)Fe$_2$As$_2$, Ref.~\onlinecite{Singleton}, ($T_c$=28~K, $T_h$=20~K, using different criteria for resistive transition, zero resistance- black triangles, midpoint- red circles, onset - green squares); (f) KFe$_2$As$_2$, Ref.~\onlinecite{Terashima}, ($T_c$=3.8~K, $T_h$=0.5~K).
}%
\label{othermaterials}
\end{figure*}

To check if Eq.~\ref{Hc2-theta} describes our data, instead of commonly used data fitting, as shown in the bottom panels of Figs.~\ref{angleslight} and \ref{anglestrong}, we used an approach based on data transformation so as to make possible deviations clearly visible.  According to Eq.~\ref{Hc2-theta}, the $H_{c2}^{-2}$ vs ($\sin^2\theta)$ should be a straight line, and in Figs.~\ref{angleslightanalysis} and \ref{anglestronganalysis} we plot this way the data for samples with $x=$0.054 and $x=$0.72 respectively. The data show clear deviation from linear trend, irrespectively of the criterion of $H_{c2}$ determination from the resistivity data, with the deviation being the strongest close to $H \parallel c-$axis or $\sin^2\theta$=1.
To check if the deviation from Eq.~\ref{Hc2-theta} in Figs.~\ref{angleslightanalysis} and \ref{anglestronganalysis} can be caused by finite inclination angle $\varphi$ (see Fig.~\ref{scheme} for the definition), here we provide the angular dependence of $H_{c2}$ for arbitrary $\varphi$
Choosing  the cross-section of the plane, in which $\bm H$ is rotated, with the $ab$ crystal plane (see Fig.~\ref{scheme}) as the $x$ axis, we obtain in the crystal frame ${\hat c}=(0,0,1)$ and the unit vector along the field ${\hat h}=(\cos\theta,\, \sin\theta\sin\varphi,\,\sin\theta\cos\varphi)$. This gives for the angle $\theta_c$ between the field and $c$ axis: $\cos\theta_c={\hat c}\cdot {\hat h}=\sin\theta\cos\varphi$. We then obtain for geometry of our experiment:

\begin{equation}
H_{c2}(\theta,\varphi)=\frac{H_{c2,ab}}{\sqrt{(\gamma_H^2-1)\cos^2\varphi \,\sin^2 \theta +1} }\,.
\label{eq2}
\end{equation}

It is seen that constant $\varphi$, as determined by our experimental geometry, does not change the linear relation of $H_{c2}^{-2}$ vs $\sin^2\theta$, despite changing the magnitude of the variation, vanishing for $\varphi=90^o$, corresponding to field rotation parallel to the conducting plane. Therefore the linear dependence of $H_{c2}^{-2}$ on $\sin^2\theta$ is not affected by a misalignment $\varphi$.

The $H_{c2}(\theta)$ described by the Eq.~\ref{Hc2-theta}, is a direct consequence of the linearized GL equation for anisotropic materials at $H_{c2}$:

\begin{equation}
- (\xi^2)_{ik}  \Pi_i\Pi_k \Psi =\Psi\,,
\label{GL}
\end{equation}

\noindent
where ${\bm  \Pi} =\nabla +2\pi i{\bm A}/\phi_0$, ${\bm  A}$ is the vector potential and $\phi_0$ is the flux quantum; summation is implied over repeating indices. Both sides of this equation are scalars, so that $(\xi^2)_{ik}$ is a second rank tensor with the standard angular dependence which is reflected in Eq.~(\ref{Hc2-theta}).

We note that in the original papers, the angular dependence, Eq.(\ref{Hc2-theta}), has been derived for single band s-wave superconductors.
It has also been recently shown that this behavior is expected for arbitrary Fermi surface, the superconducting gap modulation and for multi - band materials \cite{KoganProzorov}. However, this conclusion is achieved assuming the explicit factorization of the pairing potential and order parameter, $V(\bm k,\bm k^\prime)=V_0\Omega(\bm k)\Omega(\bm k^\prime)$ and $\Delta(T,k) = \Psi({\bf r},T)\,\Omega({\bf k}_F)$, see detailed discussion in the theory section below. There is no microscopic justification for such factorization in complex superconductors and deviations from Eq.(\ref{Hc2-theta}) can be naturally explained by violation of this procedure. In addition, for iron-pnictides the importance of the paramagnetic effects for magnetic fields parallel to the Fe-As plane was suggested to explain unusual shape of the $H_{c2}(T)$ \cite{Terashima,Singleton,LiFeAs_Kyuil}). This may also lead to the deviation from Eq.(\ref{Hc2-theta}) with the maximum effect expected at low temperatures and for orientations close to $H \parallel ab-$planes.

Motivated by these considerations, we compile in Fig.~\ref{othermaterials} the published data for various layered materials, analyzed by plotting $H_{c2}^{-2}$ vs $\sin^2\theta$. The data are arranged with decreasing anisotropy from top to bottom.
The most anisotropic materials, staged graphite intercalation compounds (top panel, data from Ref.~\onlinecite{Iye}) and layered Sr$_2$RuO$_4$ (data from Ref.~\onlinecite{Deguchi}) closely follow Eq.~\ref{Hc2-theta}. Interestingly, that clear deviations from this behavior in Sr$_2$RuO$_4$, arising due to unusual limiting mechanism in magnetic fields close to $H \parallel ab$  \cite{Deguchi}, is very difficult to recognize in a limited angular range near $\theta$=0, as the dependence in the whole range is dominated by the anisotropy of the Fermi surface. On the other hand, two materials in which superconductivity shows strong multiband features, MgB$_2$ \cite{MgB2Al} and NbSe$_2$ \cite{NbSe2} show distinctly different angular dependences. The $H_{c2}^{-2}(\sin \theta)$ in pure MgB$_2$ \cite{Tamegai} shows downward bent as field approaches $c-$ axis, $\theta$=90$^{\circ}$, similar but much less pronounced than in our observations in BaNi122. On the other hand, doped Mg(B$_{1-x}$Al$_x$)$_2$ closely follows the linear $H_{c2}^{-2}(\sin \theta)$ dependence, Eq.~\ref{Hc2-theta}, which may be suggestive that doping diminishes multi-band effects due to enhanced inter-band scattering. For pure NbSe$_2$ the $H_{c2}^{-2}(\sin \theta)$ plot shows most clear deviations from linearity among all materials, with an upward curvature towards $\theta=90^{\circ}$, of opposite trend to pure MgB$_2$ and BaNi122. 
The two angular data sets for profoundly multi-band iron pnictide superconductors, slightly underdoped, BaK122 \cite{Singleton}, and heavily overdoped K122 \cite{Terashima}, generally follow linear dependence despite profound difference in the superconducting gap structure, nodeless in the former case \cite{ReidSUST} and with vertical line nodes in the latter \cite{ReidPRL}.
Considering that among all the materials for which we were able to find published $H_{c2}(\theta)$, only pure multi-band  MgB$_2$ and NbSe$_2$ reveal clear deviations from Eq.(\ref{Hc2-theta}), it is tempting to relate the observed deviations with the multi-band superconductivity in the clean limit. This might be quite natural that in these systems the factorization of the pairing potential and of the order parameter does not hold given the complexity of the in- and inter - band interactions. This explanation, however, is not universal, since multi-band effects are very pronounced in high purity crystals of KFe$_2$As$_2$, but no clear deviations from Eq.(\ref{Hc2-theta}) are found there. On the other hand, it is hard to consider overdoped BaNi122 as a clean system, since scattering due to substitutional disorder, especially on the Fe site, is significant in these compositions. The observation that the deviations from the linear plot in MgB$_2$ diminish with disorder, suggest that it is $k$-dependence of the gap magnitude, rather than multi-band nature of the Fermi surface itself is important for the unusual angular dependence. This conclusion is in line with the recent extension of the HW theory for multiband superconductors with arbitrary Fermi surfaces \cite{KoganProzorov}.

In discussing these results we should keep in mind, that in all cases, except for Sr$_2$RuO$_4$, the $H_{c2}$ was measured resistively, so that inevitably its determination is approximate since the resistive transition as a rule has finite width and hence, the $H_{c2}$ values depend on a criterion chosen. Finite resistivity in the flux - flow regime (most pronounced in the clean systems) broadens the transition making resistive determination difficult. From this point of view, assertions of Kim {\it et al.} \cite{MgB2Al} that their data allow to distinguish between two models, GL and two - band Usadel approach by Gurevich \cite{MgB2theta}, are hard to accept.

In compounds with relatively high $T_c$, the determination of $H_{c2}$ from resistive measurements is also complicated by the phenomenon of vortex lattice melting: above the melting point, the resistivity is close to that of the normal phase and $H_{c2}$ {\it per se} becomes invisible in resistivity measurements. This complication in a given material might affect the measurements stronger near $T_c$ than at low temperatures.

\subsubsection{Theoretical considerations}

Recently, the theoretical justification for the near-universality of the angular dependence (Eq.~\ref{Hc2-theta})  has been offered in Ref.~\onlinecite{KoganProzorov} for orbital $H_{c2}(\theta)$ albeit under certain restricting assumptions. We reproduce briefly the argument of this work. Some deviations from the effective mass model have been predicted theoretically for quasi-one-dimensional conductors , see Ref.~\onlinecite{Lebed}. We also note that the theory of $H_{c2}(\theta)$ at the arbitrary angle with respect to the crystal axes in \textit{multi-band superconductors} has additional complications. In particular, if $H$ is not parallel to the $c-$axis and the mass anisotropy ratios in two bands are different, and the first Landau level solutions which are then substituted into the HW integrals cannot simultaneously satisfy the two - band Eilenberger equations and the gap equations, and the higher Landau levels become important \cite{Gurevich,Gurevich2010}. Perhaps this provides an alternative explanation of why we observe the deviations from the conventional scaling only in clean two - band systems \cite{Gurevich2012}. Keeping this in mind, we proceed with the generalized description \cite{KoganProzorov}.

We recall that, as HW \cite{HW} had shown within the weak-coupling BCS-Gor'kov theory, the linearized GL equation $-  \xi^2  \Pi^2 \Delta =\Delta  $ for the isotropic case holds at any $T$ at $H_{c2}(T)$.  However $\xi(T)$ does not diverge for $T\to T_c(H)<T_c(0)$, -instead it is finite and should be evaluated to satisfy the basic self-consistency (or gap) equation for the order parameter $\Delta$. It is easily shown then that $H_{c2}(T)=\phi_0/2\pi\xi^2$.

For materials with anisotropic Fermi surfaces and order parameters one can consider Eq.\,(\ref{GL}) as an ansatz for satisfying the self-consistency equation of the theory. To facilitate this derivation one assumes the coupling potential of the form  $V(\bm k,\bm k^\prime)=V_0\Omega(\bm k)\Omega(\bm k^\prime)$ and the order parameter $\Delta = \Psi({\bf r},T)\,\Omega({\bf k}_F)$. The function  $ \Omega({\bf k}_F)$ which determines the $\bm k_F$ dependence of $\Delta$ is normalized so that
$\langle\Omega^2\rangle=1$.  This convenient approximation has been first designed by  Markovic and Kadanov \cite{Kad} and it works well for one band materials with anisotropic coupling.

For two-band materials one can use a generalization of this procedure.
Other than s-wave order parameters emerge if the coupling $V(\bm k,\bm k^\prime)$ responsible for superconductivity is not   a constant (or a 2$\times$2 matrix of $\bm k$ independent constants for two bands). The formally simplest way to consider different from s-wave order parameters without going to  details of microscopic interactions is to use a ``separable" potential, $V_{\alpha\beta}(\bm k,\bm k^\prime) = V^{(0)}_{\alpha\beta}\Omega_\alpha(\bm k)\Omega_\beta(\bm k^\prime)$ where $V^{(0)}_{\alpha\beta}$ is a $\bm k$ independent matrix,
and look for the order parameters in the form
$ \Delta_{\alpha } (\bm r,T,\bm k) = \Psi_{\alpha }(\bm r,T)\Omega_\alpha(\bm k) $
with the normalization $\big\langle  \Omega ^2    \big\rangle_\alpha=1$ for each band ($\alpha,\beta=1,2$ are band indices).

It has been shown in Ref.~\onlinecite{KoganProzorov} that in this approximation the ansatz
\begin{equation}
- (\xi^2)_{ik} \Pi_i\Pi_k \Delta_\alpha=\Delta_\alpha,\qquad  \alpha=1,2 \,. \label{ansatz1}
\end{equation}
leads to solutions satisfying the self-consistency equations, which provide
all components of the tensor $ (\xi^2)_{ik}(T)$.   In this way the ansatz (\ref{ansatz1}) is proven correct. Along with $ (\xi^2)_{ik}(T)$, the upper critical field and its anisotropy can be evaluated for a  given Fermi surface and the order parameter symmetries $\Omega_\alpha$.

Since $(\xi^2)_{ik}$ is a second rank tensor, we conclude, exactly as in the case of anisotropic GL Eq.\,(\ref{GL}), that the angular dependence of Eq.\,(\ref{Hc2-theta}) should hold at any temperature and for any anisotropic Fermi surface, any symmetry of the bands order parameters, and at any temperature. In other words, within the approximation of ``separable" coupling potential, the dependence (\ref{Hc2-theta}) is ``universal".

Clearly, ``separable" potentials do not exhaust all possible interactions and, therefore, other forms of the angular dependence $H_{c2}(\theta)$ can exist. An example of such a potential has been studied in Ref.~\onlinecite{Makids}.
Such potentials may lead to  gradient terms in GL equations  different from the standard forms (\ref{GL}) or (\ref{ansatz1}) and, therefore, different from Eq.\,(\ref{Hc2-theta}) angular dependencies, see e.g. Ref.\,\onlinecite{Gork-heavy}.

We therefore may conclude that deviations of the observed angular dependence of $H_{c2}$ from the form (\ref{Hc2-theta}) (or deviations of $H_{c2}^{-2}$ plotted  {\it vs.} $\sin^2 \theta $ from the straight line) signal that the coupling potential cannot be written in the separable form. On the other hand, the example of separable potentials (for any Fermi surface and any order parameter symmetry) shows that there is no direct relation between the angular dependence of $H_{c2}$, Fermi surfaces, and order parameter symmetries. However, deviations of  $H_{c2}(\theta)$ from the form (\ref{Hc2-theta}) may carry such an information. To investigate this  question   further one would need a better data on these deviations, in particular, criterion-independent determination of $H_{c2}$, which is hard to achieve in resistive measurements. On the theoretical side, of course, one should go beyond the weak coupling and separable coupling potentials.

\subsection{Temperature dependence of $H_{c2}$ }

There are two mechanisms that determine the upper critical field of superconductors. The first one, determined by the supercurrent flow to screen the magnetic field, is referred to as orbital limiting and described by HW theory \cite{HW,WHH}. The upper critical field at $T \to 0$ limit, $H_{c2}(0)$, in HW theory is determined by the slope of the $H_{c2}(T)$ curve close to $T_c$, and as $T$ goes to zero the curve shows downward deviation from linear dependence and eventual saturation towards the value $H_{c2}(0) \approx 0.7T_c \frac{dH_{c2}}{dT}$ in isotropic case.

Rather rare exceptions, when the upper critical field is not determined by the orbital motion, are found in the materials in which orbital motion of electrons is hampered by either short mean free path, heavy mass of conduction electrons in heavy fermion materials or weak links between the conducting layers in Josephson structures or in naturally layered materials, provided that the magnetic field is aligned precisely parallel to the conducting layer. In this situation the upper critical field $H_{c2}$ is determined by Zeeman splitting of electron levels, known as Clogston-Chandrasekhar \cite{CC} paramagnetic limit. This field is determined by a decrease of paramagnetic energy becoming equal to condensation energy of superconductor. In weak coupling BCS superconductors the paramagnetic limiting field is determined in $T \to 0$ limit as $H_p$=1.84$T_c$, where $H_p$ is field in Tesla and $T_c$ is in K.

As can be seen from Figs.~\ref{phasedslight} and \ref{phasedstrong} the upper critical fields in $H \parallel ab$ configuration are higher than the weak-limit paramagnetic limiting $H_p$, equal to 32.2~T ($x=$0.054) and 13.8~T ($x=$0.072). These high values may come from strong coupling nature of superconductivity in iron pnictides, or indeed reflect paramagnetic limiting at low temperatures as was suggested in several studies \cite{Singleton,FeSeBalakirev,LiFeAs_Kyuil}.

\section{Conclusions}

By performing high angular resolution study of the upper critical field in two overdoped compositions of iron pnictide superconductor Ba(Fe$_{1-x}$Ni$_x$)$_2$As$_2$, we find clear deviations from the anisotropic Ginzburg-Landau form. Implementing linearization plot analysis of our and previously published data, we find clear deviations from the form only in the case of clean fully - gapped multi-band superconductors, such as NbSe$_2$ ang MgB$_2$, but not in dirty MgB$_2$ and not clean KFe$_2$As$_2$. We speculate, that the dependence may reflect $c$-axis modulation of the superconducting gap, as suggested by anisotropic penetration depth and thermal conductivity measurements \cite{Tanatarthermalcond,Reidcaxis,Martin}.

\section{Acknowledgements}
We thank A. Gurevich and A.~G.Lebed' for useful comments. Work at the Ames Laboratory was supported by the Department of Energy-Basic Energy Sciences under Contract No. DE-AC02-07CH11358. JSB acknowledges support from NSF DMR 1005293.
Work at the National High Magnetic Field Laboratory is supported by the NSF Cooperative Agreement No. DMR0654118 and by the state of Florida. 

%%%%%%%%%%%%%%%%%%%%%%%%%%%% BIBLIOGRAPHY

\end{document}